\documentclass[conference]{IEEEtran}
\IEEEoverridecommandlockouts

\normalsize

\ifCLASSINFOpdf
\else
\fi
\usepackage{setspace} 

\usepackage{cite}
\usepackage{ragged2e}
\usepackage{amssymb}
\usepackage{paralist}
\usepackage{amsmath}
\usepackage{amsthm}
\usepackage{graphicx}
\usepackage{epstopdf}
\usepackage{color}
\usepackage{extarrows}
\usepackage{multicol}
\usepackage{caption}
\usepackage{url}
\usepackage{bm}
\usepackage{footnote}
\usepackage{diagbox}
\usepackage{multirow}
\usepackage{array}
\usepackage{diagbox}
\usepackage{cases}

\usepackage[linesnumbered,ruled,vlined]{algorithm2e}
\usepackage{algorithmicx}

\usepackage[justification=centering]{subcaption}
\usepackage{booktabs}

\allowdisplaybreaks
\PassOptionsToPackage{bookmarks=false}{hyperref}

\def\be{ \begin{eqnarray} }
\def\ee{ \end{eqnarray} }
\usepackage{soul}
\usepackage{stfloats}

\begin{document}

\title{Meta Learning Based Adaptive Cooperative Perception in Nonstationary Vehicular Networks}





\author{\IEEEauthorblockN{1\textsuperscript{st} Kaige Qu}
\IEEEauthorblockA{\textit{Department of Electrical} \\
\textit{ and Computer Engineering,}\\
\textit{University of Waterloo}\\
Waterloo, Canada \\
k2qu@uwaterloo.ca}
\and
\IEEEauthorblockN{2\textsuperscript{nd} Zixiong Qin}
\IEEEauthorblockA{\textit{School of Electronics Engineering,} \\
\textit{Beijing University of}\\
\textit{Posts and Telecommunications}\\
Beijing, China \\
zxqin@bupt.edu.cn}
\and
\IEEEauthorblockN{3\textsuperscript{rd} Weihua Zhuang}
\IEEEauthorblockA{\textit{Department of Electrical} \\
\textit{ and Computer Engineering,}\\
\textit{University of Waterloo}\\
Waterloo, Canada \\
wzhuang@uwaterloo.ca}
}

\maketitle

\begin{abstract}

To accommodate high network dynamics in real-time cooperative perception (CP), reinforcement learning (RL) based adaptive CP schemes have been proposed, to allow adaptive switchings between CP and stand-alone perception modes among connected and autonomous vehicles. The traditional offline-training online-execution RL framework suffers from performance degradation under nonstationary network conditions. To achieve fast and efficient model adaptation, we formulate a set of Markov decision processes for adaptive CP decisions in each stationary local vehicular network (LVN). A meta RL solution is proposed, which trains a meta RL model that captures the general features among LVNs, thus facilitating fast model adaptation for each LVN with the meta RL model as an initial point. Simulation results show the superiority of meta RL in terms of the convergence speed without reward degradation. The impact of the customization level of meta models on the model adaptation performance has also been evaluated.

\end{abstract}

\begin{IEEEkeywords}

Meta reinforcement learning, cooperative perception, connected and autonomous vehicles (CAVs), nonstationary vehicular networks.

\end{IEEEkeywords}

\IEEEpeerreviewmaketitle

\section{Introduction}

The advances in sensing, artificial intelligence (AI), vehicles-to-everything (V2X) communication technologies, and vehicular edge computing (VEC) paradigms are revolutionizing the future transportation systems with connected and autonomous vehicles (CAVs), which will improve the road safety and traffic efficiency~\cite{Proceedings,shen2021holistic,hui2022collaboration}. Ideally, each CAV has the stand-alone perception (SP) capability for detecting the presence, locations, and classes of surrounding objects, to maintain situation awareness and support autonomous driving applications such as maneuver control and path planning~\cite{wang2018networking,zhang2019mobile}. In practice, SP is vulnerable to obstruction, distance, and poor sensor quality at bad photometric conditions. As a supplement to SP, cooperative perception (CP) has been proposed to enable the sensory information sharing among CAVs, allowing the CAVs to collaboratively perceive a more comprehensive and accurate view of the environment~\cite{zheng2022confidence,xiao2022perception,sun2022user,lin2022low}.

Cooperative perception is a hot research topic in the computer vision field, with a focus on the design of advanced data fusion models. 
Research efforts have also been made for supporting secure and trusted cooperative operations among unmanned vehicles based on blockchain technology~\cite{9743876,9446552}.
From the networking perspective, there are several challenges such as scalability, performance uncertainty, and network dynamics, that hinder accurate and real-time CP~\cite{10400178}. There have been several studies on CAV association and data selection for CP to enhance the perception accuracy with efficient network resource usage~\cite{jia2023mass,abdel2021vehicular,10363435}. In our previous work, several approaches have been explored to address the network dynamics issue, including the adaptive switching between SP and CP modes and the dynamic radio resource allocation to adapt to varying network conditions, and the dynamic voltage and frequency scaling (DVFS) at the CPUs to accommodate the perception workload and computing demand variations~\cite{10413956}. A reinforcement learning (RL) based solution is proposed in~\cite{10413956} to learn the adaptive cooperation decisions among CAV pairs under network dynamics, which integrates a resource allocation optimization model for reward calculation. 

\begin{figure*}
    \centering
    \includegraphics[width=0.8\linewidth]{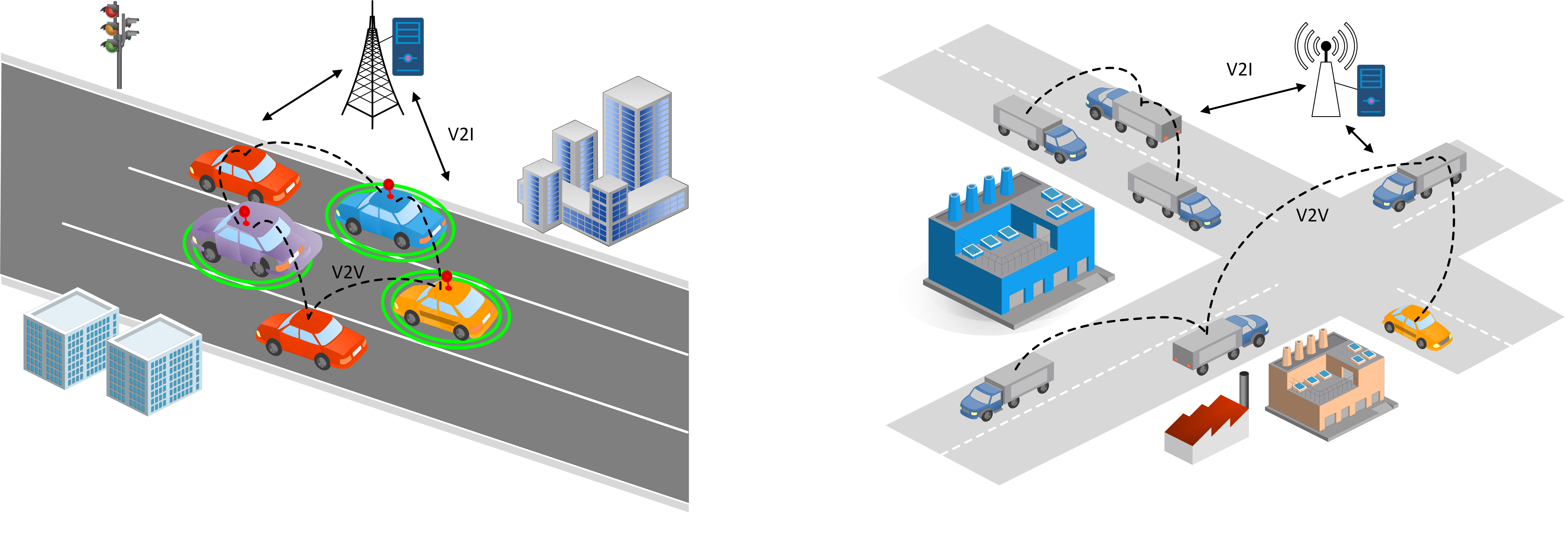} 
  \caption{Vehicular networks with spatio-temporal nonstationary network dynamics.}\label{fig:scenarios}
\end{figure*}

Typically, there are offline-training and online-execution phases in an RL solution, where an RL model is trained offline to capture the system dynamics by learning from historical experiences and used online for real-time decisions. We consider a number of local vehicular networks (LVNs) with nonstationary spatio-temporal characteristics in a geographical region, as illustrated in Fig.~\ref{fig:scenarios}. For example, the network conditions in different LVNs at different time periods can have significant changes. In such cases, the offline-training and online-execution RL framework performs poorly, as the learned knowledge becomes invalid under new conditions.

To enhance the robustness of machine learning based network optimization solutions, fast and efficient learning model adaptation techniques under a nonstationary network condition need further investigation. Retraining from scratch is time-consuming and data-inefficient~\cite{benzaid2020ai}. Retraining with an old model as the initial point shows uncertain performance, as it is difficult to estimate the similarity between the unknown old and new underlying distributions of network dynamics. The performance is deteriorated if the new distribution significantly deviates from the old one. Meta learning is a “learning to learn” technique that aims at learning how to efficiently execute new learning tasks. During a meta training phase, a meta model with good generalization ability is trained by learning from a diverse set of individual learning tasks, to learn a high-level initialization that quickly adapts to new tasks. Then, during a meta adaptation phase, the trained meta model serves as an initial point for fast model adaptation on new learning tasks with a limited amount of data.

In this paper, we first formulate a joint adaptive CAV cooperation and resource allocation problem based on our previous work to address the network dynamics in cooperative perception~\cite{10413956}. The objective is to maximize the total computing efficiency gain while minimizing the total switching cost. Then, we reformulate the problem to a set of Markov decision processes (MDPs) with the consideration of the spatio-temporal nonstationarity among LVNs. A meta RL solution which integrates the meta learning algorithm and the proximal policy optimization (PPO) algorithm, referred to as meta-PPO algorithm, is proposed to solve the set of MDPs. Each MDP is associated with an individual PPO model that learns the adaptive perception mode selection decision sequences among multiple CAV pairs, under an unknown but stationary distribution of network dynamics. Instead of directly training the individual PPO models for all the MDPs, a meta PPO model is trained to capture the general features among all the MDPs, which then serves as an initial point for training individual PPO models via fast model adaptation.

The remainder of this paper is organized as follows. The system model is presented in Section II. The joint adaptive CAV cooperation and resource allocation problem is formulated in Section III, which is reformulated to a set of MDPs in Section IV for nonstationary networks. A meta RL solution is presented in Section V, and simulation results are provided in Section VI, followed by conclusions in Section VII.

\section{System Model}

\subsection{Network Scenario}

We consider a vehicle cluster including $K$ CAV pairs in set $\mathcal{K}$, under the service coverage of a base station or a road side unit. The CAV pairs are predetermined by using existing CAV association algorithms [11], [12]. Both CAVs in a pair share a similar view but from different angles. Consider a time-slotted system, in which the time slots are indexed by integer $t$. In each time slot, a CAV initiates a perception task to identify and classify the surrounding objects. To support the perception tasks, each CAV pair is allowed to work in an SP mode by default and in a CP mode by selection. 
Let $\boldsymbol{a}_t=\left\{ x_k(t), \forall k\in\mathcal{K}\right\}$ be a binary perception mode selection decision vector for all CAV pairs at time slot $n$, with $x_k(t)=1$ indicating the CP mode and $x_k(t)=0$ indicating the SP mode for CAV pair $k$.
A CAV pair in the CP mode is referred to as a cooperative CAV pair. 
Let $\mathcal{K}_C(t)$ denote a set of cooperative CAV pairs at time slot $n$, with $\mathcal{K}_C(t) \subset \mathcal{K}$.

\subsection{Perception Task Model}

For both CAVs in a pair, the objects in their overlapping and non-overlapping sensing ranges are referred to as shared and individual objects respectively. Let $W_k(t)$ denote the number of shared objects, also referred to as shared workload, for CAV pair $k$. In the SP mode, both CAVs independently perform all the object classification tasks by using a locally deployed default DNN model. For a cooperative CAV pair, the shared objects are collaboratively processed by using a feature-fusion DNN model which is partitioned between the transmitter and receiver CAVs, while the remaining individual objects are processed by using the default DNN models~\cite{10413956}. Specifically, for each shared object, both CAVs extract features from the local sensing data. Then, the feature data of the transmitter CAV are sent to the receiver CAV via V2V communication, where the feature data from both CAVs are fused and processed by deeper DNN layers. The object classification result obtained at the receiver CAV is then returned to the transmitter CAV, with negligible transmission delay due to the small data size. Due to the data fusion and further processing at only the receiver CAV, there is a positive total computing demand reduction for a cooperative CAV pair $k$, given by
\begin{align} 
A_k(t) = \left(2\delta - \tilde{\delta}\right) W_k(t)	
\end{align}
where $\delta$ and $\tilde{\delta}$ represent the average computing demand (in CPU cycles) for processing one object by using the default and feature-fusion DNN models, respectively~\cite{10413956}.

To support the dynamic perception workloads, a dynamic voltage and frequency scaling (DVFS) technique is used at the CAVs to allow on-demand CPU frequency scaling. For simplicity, assume that the shared and individual objects are processed at separate CPU cores. Let $\boldsymbol{f}(t)=\left\{ f_k(t), \forall k\in\mathcal{K}\right\}$ be a CPU frequency allocation decision vector for processing the shared objects at all CAV pairs at time slot $t$. 
For CAV pair $k$, we have $f_k(t) = \frac{  \delta W_k(t)}{\Delta}$ in the SP mode, to ensure that the default DNN model processing can be finished within a
delay bound, $\Delta$, for the shared objects. We use a separate parameter, $f^{\mathsf{D}}_k(t)$, to represent the default CPU frequency
for processing the shared objects in the SP mode, given by $f^{\mathsf{D}}_k(t) = \frac{  \delta W_k(t)}{\Delta}$. The relationship among $f_k(t)$, $f^{\mathsf{D}}_k(t)$, and $x_k(t)$ can be expressed as
\begin{align} 
&\left(1-x_k(t)\right)f^{\mathsf{D}}_k(t) \leq f_k(t) \nonumber\\
&\leq \left(1-x_k(t)\right)f^{\mathsf{D}}_k(t) + x_k(t)\mathbb{M}, \quad \forall k\in\mathcal{K}	
\label{eq-CPU-relation} 
\end{align}
where $\mathbb{M}$ is a very large constant.

To evaluate the computing efficiency for processing the perception tasks, we define a computing efficiency gain of
a CAV pair as the reduced amount of computing energy in comparison with that in the default SP mode, denoted by $G_k(t)$ for CAV pair $k$ at time slot $t$. Such a performance metric evaluates the impact of both the CPU frequency and the computing demand on the computing efficiency. According to~\cite{10413956}, we have
\be 
	G_k(t) =  \left\{ 
	\begin{array}{rcl}
	0, \hspace{0mm}     & {\forall k\notin\mathcal{K}_C(t)}\\
	\kappa W_k(t) \left[ 2\delta f^{\mathsf{D}}_k(t)^2  - \tilde{\delta} f_k(t)^2 \right],  & {\forall k\in\mathcal{K}_C(t)}
	\end{array} \right.
\ee
where $\kappa$ is the energy efficiency coefficient of a CPU core. 
Here, $G_k(t)$ is a decreasing function of $f_k(t)>0$. For CAV
pair $k\in\mathcal{K}_C(t)$, $G_k(t)$ can be negative if $f_k(t)$ is greater than a parameter which depends on the DNN model structures, the shared workload, and the delay bound~\cite{10413956}. Let $f^{\mathsf{P}}_k(t)$ denote such a parameter, which is expressed as
\begin{align} 
    f^{\mathsf{P}}_k(t) = \sqrt{\frac{2\delta^3}{\tilde{\delta}}} \frac{W_k(t)}{\Delta}.
\end{align}
Let $f_{\mathsf{M}}$ denote the maximum CPU frequency supported by DVFS. Then, for each cooperative CAV pair, the following constraint should be satisfied:
\begin{align} 
    f_k(t) \leq \min \left\{ f^{\mathsf{P}}_k(t),  f_{\mathsf{M}} \right\}, \quad \forall k\in\mathcal{K}_C(t) 
    \label{eq-CPU-freq-ideal}
\end{align}
to ensure $G_k(t) \geq 0$ at a feasible CPU frequency.

\subsection{Communication Model}

Let $B(t)$ be the available radio spectrum bandwidth for CAVs at time slot $t$, with the consideration of dynamic background
radio resource usage by other vehicle users. Consider orthogonal frequency division multiplexing (OFDM) based
transmission schemes for the CAVs.

Let $\boldsymbol{\beta}(t)=\left\{ \beta_k(t), \forall k\in\mathcal{K}\right\}$ be a radio resource allocation decision vector for all CAV pairs at time slot $t$, with $\beta_k(t)$ denoting the fraction of available radio resources allocated to CAV pair $k$. To guarantee that no radio resources are allocated to CAV pairs in the SP mode, we should have
\begin{align} 
    0 \leq \beta_k(t) \leq x_k(t), \quad \forall k\in\mathcal{K}.
    \label{eq-radio-relation} 
\end{align}
To ensure that the total fraction of allocated bandwidth for all CAV pairs does not exceed one, we should have
\begin{align} 
    \sum_{k\in\mathcal{K}} \beta_k(t) \leq 1. 
    \label{eq-total-radio}
\end{align}
The average transmission rate for CAV pair $k$ at time slot $t$ is
\begin{align} 
    R_k(t) =  \beta_k(t) B(t) \log_2 \left( 1+\frac{ p_k g_k(t) }{ \sigma^2} \right), \quad \forall k\in\mathcal{K}
    \label{eq-achieved-rate}
\end{align}
where $p_k$ is the transmit power of CAV pair $k$, $g_k(t)$ is the channel power gain between both CAVs in CAV pair $k$ at time slot $t$, and $\sigma^2$ represents the received noise power. 

\section{Optimization Problem Formulation}

At a cooperative CAV pair, the feature extraction, feature data transmission, feature fusion, and fused data processing for each shared object should be completed within an average delay budget, $\frac{\Delta}{W_k(t)}$. Such a delay constraint depends on the radio resource and CPU frequency allocation decisions, represented as $\frac{w}{R_k(t)} + \frac{ \hat{\delta}}{f_k(t)} \leq \frac{\Delta}{W_k(t)}, \forall k\in\mathcal{K}_C(t)$,
where $w$ is the feature data size and $\hat{\delta}$ is a DNN model structure dependent constant~\cite{10413956}. We have proved that the
delay constraint should be active to avoid resource overprovisioning, rewritten as
\begin{align} 
    \frac{w}{R_k(t)} + \frac{ \hat{\delta}}{f_k(t)} = \frac{\Delta}{W_k(t)}, \quad \forall k\in\mathcal{K}_C(t)
    \label{eq-delay-equality}
\end{align}

Under the network dynamics in terms of radio resource availability, perception workloads and inter-vehicle distances, a subset of CAV pairs should be selected to work in the CP mode at each time slot, to maximize the total computing efficiency gain. Moreover, the number of switchings between the SP and CP modes among all the CAV pairs should be minimized, to reduce the CPU process switching overhead between scheduling the default and feature-fusion DNN models.
Let $C(t)$ be the total number of CAV pairs that change the cooperation status at time slot $t$, given by
\begin{align} 
	C(t) = \sum_{k\in\mathcal{K}} \left| x_k(t) - x_k(t-1) \right|. 
    \label{eq-switch-cost} 
\end{align} 
The total switching cost increases proportionally to $C(t)$.
Therefore, to maximize the total computing efficiency gain while minimizing the total switching cost in the long run, we formulate a joint perception mode selection, radio resource allocation, and CPU frequency allocation problem,
\begin{align} 
    \mathbf{P}_0: ~\max_{ \boldsymbol{a},\boldsymbol{\beta},\boldsymbol{f} }   \hspace{0.3cm} &{ \sum_{t} \left[\left(\sum_{k\in\mathcal{K}}G_{k}(t)\right) - \tilde{\omega} C(t)\right] }    \label{eq-obj-joint}\\
    \text{s.t. } \hspace{0.3cm}  &  \eqref{eq-CPU-relation}, \eqref{eq-CPU-freq-ideal}, \eqref{eq-radio-relation}, \eqref{eq-total-radio}, \eqref{eq-delay-equality} \nonumber
\end{align}
where $\tilde{\omega}$ is a positive weight that controls the trade-off between the computing efficiency gain and switching cost.

\section{Markov Decision Process Reformulation for Nonstationary Networks}

With the consideration of the spatio-temporal nonstationarity in vehicular networks, we reformulate problem $\mathbf{P}_0$ to a set of MDPs. Each MDP is associated with one LVN with a stationary underlying distribution of the network dynamics.
For LVNs with different statistical characteristics, consider an unknown distribution, $\rho(\mathcal{T})$, over a set $\mathcal{T}$ of MDP tasks which share the state space, action space, and reward function but differ in the state transition probabilities. 

An MDP task, $\mathcal{T}_i \sim \rho(\mathcal{T})$, which is associated with an LVN, is represented as a tuple, $\left(\mathcal{S}, \mathcal{A}, P_i, R, \gamma\right)$, where $\mathcal{S}$ is the state space, $\mathcal{A}$ is the action space, $P_i$ is the task-specific state transition probability matrix, with $P_i(\boldsymbol{s}'|\boldsymbol{s},\boldsymbol{a})$ denoting the probability distribution of next state $\boldsymbol{s}'\in\mathcal{S}$ given current state $\boldsymbol{s}\in\mathcal{S}$ and action $\boldsymbol{a}\in\mathcal{A}$, 
$R: \mathcal{S}\times\mathcal{A}\mapsto \mathbb{R}$ is a reward function, 
and $\gamma$ is a discount factor in $[0,1)$. 
Let $\tau_i = \{\boldsymbol{s}_1, \boldsymbol{a}_1, r_1, \cdots, \boldsymbol{s}_T, \boldsymbol{a}_T, r_T\}$ denote a trajectory of MDP task $\mathcal{T}_i$, which collects the state, action, and reward for $T$ time steps. 
In our MDP for adaptive cooperative perception, the state, action, and reward are given as follows.
\begin{itemize}

	\item \emph{State}: The state at time step $t$, denoted by $\boldsymbol{s}_t$, includes the available radio spectrum bandwidth for CAVs, $B(t)$, the shared workloads of all CAV pairs, $\left\{W_k(t),\forall k\in\mathcal{K}\right\}$, the transmitter-receiver distances of all CAV pairs, $\left\{D_k(t),\forall k\in\mathcal{K}\right\}$, and the previous cooperation status of all CAV pairs, $\left\{x_k(t-1),\forall k\in\mathcal{K}\right\}$, given by
	\begin{align} 
		\boldsymbol{s}_t = \left\{ B(t), W_k(t), D_k(t), x_k(t-1),\forall k\in\mathcal{K} \right\};
	    \label{eq-original-obs} 
	\end{align}

	\item \emph{Action}: The action at time step $t$ is the binary perception mode selection decision vector, $\boldsymbol{a}_t=\{x_k(t),\forall k\in\mathcal{K}\}$;

	\item \emph{Reward}: 
	For time slot $t$, given action $\boldsymbol{a}_t$, a cooperative CAV pair set, $\mathcal{K}_C(t)$, is determined, and $\mathbf{P}_0$ is reduced to a resource allocation subproblem, given by
	\begin{align} 
	    \mathbf{P}_1: ~\max_{ \boldsymbol{\beta}_C(t), \boldsymbol{f}_C(t) }   \hspace{0.3cm} &{ \sum_{k\in\mathcal{K}_C(t)}G_{k}(t) }  \\
	    \text{s.t. } \hspace{0.3cm}  &  \eqref{eq-CPU-relation}, \eqref{eq-CPU-freq-ideal}, \eqref{eq-radio-relation}, \eqref{eq-total-radio}, \eqref{eq-delay-equality} 
	\end{align}
	where $\boldsymbol{\beta}_C(t)=\{\beta_k(t),\forall k\in\mathcal{K}_C(t)\}$ and $\boldsymbol{f}_C(t)=\{f_k(t),\forall k\in\mathcal{K}_C(t)\}$ are the resource allocation decisions for CAV pairs in $\mathcal{K}_C(t)$.  
	For CAV pair $k$ in the SP mode, we have $\beta_k(t)=0$, $f_k(t)=f_k^{\mathsf{D}}(t)$, and $G_k(t)=0$. Hence, if problem $\mathbf{P}_1$ is feasible, we have $G^\ast(t)=\sum_{k\in\mathcal{K}_C(t)}G_{k}^\ast(t)$ as the maximal total computing efficiency gain with optimal resource allocation; otherwise, $G^\ast(t)$ is undefined and a negative penalty, $\dot{\mathsf{P}}$, should be applied. Therefore, the reward at time step $t$, denoted by $r_t$, is given by
	\be 
	r_t =  \left\{ 
	\begin{array}{ll}
	G^\ast(t) -  \tilde{\omega}C(t)   \text{,}      & { \mbox{if $\mathbf{P}_1$ is feasible} }\\
	\dot{\mathsf{P}}  \text{,}     & { \mbox{if $\mathbf{P}_1$ is infeasible}. } 
	\label{eq-sys-reward}
	\end{array} \right.
	\ee
	We have derived the optimal solution to problem $\mathbf{P}_1$ based on Karush-Kuhn-Tucker (KKT) conditions~\cite{10413956}. Details are neglected here.

\end{itemize}

\section{Meta Reinforcement Learning Solution}
\label{sec:Meta Reinforcement Learning Solution}

We propose a meta RL solution which integrates meta learning and RL, to train a meta RL model with a good generalization ability among different LVNs with spatio-temporal nonstationary characteristics. The meta RL model learns the general features among all MDP tasks in $\rho(\mathcal{T})$. Once the meta RL model is trained, an individual RL model for any unseen MDP from $\rho(\mathcal{T})$ can be adapted from it as an initial point, facilitating fast model adaptation under nonstationary network conditions. Here, we use PPO as the RL algorithm.

\subsection{Background of PPO Algorithm}

A PPO learning agent trains two deep neural networks: a critic network, $V_{\boldsymbol{w}}(\boldsymbol{s})$, parameterized by weights $\boldsymbol{w}$, and an
actor network, $\pi_{\boldsymbol{\phi}}(\boldsymbol{a}|\boldsymbol{s})$, parameterized by weights $\boldsymbol{\phi}$~\cite{schulman2017proximal}.
The critic network approximates a value function, $V(\boldsymbol{s}_t) = \mathbb{E}\left[\sum_{t'=t}^T \gamma^{t'-t} r_{t'} \big| \boldsymbol{s}_t  \right]$, which evaluates the expected cumulative discounted rewards that can be obtained from being in state $\boldsymbol{s}_t$.   
The actor network trains a stochastic policy function that maps state $s$ to a categorical distribution over actions in a discrete action space. 

For sampling efficiency, PPO leverages off-policy data for training via importance sampling. 
In one training iteration, the learning agent uses a sampling policy, $\pi_{\boldsymbol{\phi}^{old}}(\boldsymbol{a}|\boldsymbol{s})$, with old actor network weights $\boldsymbol{\phi}^{old}$ to interact with the environment and to sample trajectory data.
Let $\mathcal{D}=\{\tau = \{\boldsymbol{s}_1, \boldsymbol{a}_1, r_1, \cdots, \boldsymbol{s}_T, \boldsymbol{a}_T, r_T\}\}$ denote a set of trajectories collected by using the sampling policy. 
A target policy, $\pi_{\boldsymbol{\phi}}(\boldsymbol{a}|\boldsymbol{s})$, is learned by training the actor network based on $\mathcal{D}$, through maximizing a clipped surrogate objective function, $\mathcal{J}^a(\boldsymbol{\phi})$, via gradient ascent on $\boldsymbol{\phi}$. We have 
\begin{align}  
    \mathcal{J}^a(\boldsymbol{\phi}) = \mathbb{E}_{\tau\in\mathcal{D},\left(\boldsymbol{s}_t,\boldsymbol{a}_t, r_t\right) \in \tau}  \left\{ \min \left[ \frac{\pi_{\boldsymbol{\phi}}(\boldsymbol{a}_t|\boldsymbol{s}_t)}{\pi_{\boldsymbol{\phi}^{old}}(\boldsymbol{a}_t|\boldsymbol{s}_t)}  \hat{A}_t , \right.\right. \nonumber\\
    \left.\left. \texttt{clip}\left( \frac{\pi_{\boldsymbol{\phi}}(\boldsymbol{a}_t|\boldsymbol{s}_t)}{\pi_{\boldsymbol{\phi}^{old}}(\boldsymbol{a}_t|\boldsymbol{s}_t)}, 1-\epsilon, 1+\epsilon \right) \hat{A}_t  \right] \right\}
    \label{eq-obj-clip}
\end{align}
%
and the gradient ascent step as
\begin{align}
	\boldsymbol{\phi} \leftarrow \boldsymbol{\phi} + \alpha_a \nabla_{\boldsymbol{\phi}} \mathcal{J}^a(\boldsymbol{\phi})
	\label{eq-actor-update}
\end{align}
where $\mathbb{E}_{\tau\in\mathcal{D},\left(\boldsymbol{s}_t,\boldsymbol{a}_t, r_t\right) \in \tau}$ denotes the empirical average over all time steps in sampled trajectory set $\mathcal{D}$, $\hat{A}_t$ is the estimated advantage value of taking action $\boldsymbol{a}_t$ at state $\boldsymbol{s}_t$, $\epsilon$ is a clipping hyperparameter, and $\alpha_a$ is the actor learning rate. 
In each training iteration, a gradient descent is performed on the target policy, $\pi_{\boldsymbol{\phi}}\left(\boldsymbol{a}|\boldsymbol{s}\right)$. Over every multiple training iterations, the sampling policy, $\pi_{\boldsymbol{\phi}^{old}}\left(\boldsymbol{a}|\boldsymbol{s}\right)$, is replaced by the updated target policy, and used for data sampling.

The critic network is trained by minimizing the mean squared error between the estimated value, $V_{\boldsymbol{w}}(\boldsymbol{s}_t)$, and the cumulative discounted reward, $\sum_{t'=t}^T \gamma^{t'-t} r_{t'}$, at all time steps in the sampled trajectories. Accordingly, the loss function for training the critic network is 
\begin{align}  
    \mathcal{L}^c(\boldsymbol{w}) = \mathbb{E}_{\tau\in\mathcal{D},\left(\boldsymbol{s}_t,\boldsymbol{a}_t, r_t\right) \in \tau} \left(  V_{\boldsymbol{w}}(\boldsymbol{s}_t)  - \sum_{t'=t}^T \gamma^{t'-t} r_{t'} \right)^2
    \label{eq-critic-loss}
\end{align}
based on which weights $\boldsymbol{w}$ are updated via a gradient descent in each training step, represented as
\begin{align}
	\boldsymbol{w} \leftarrow \boldsymbol{w} - \alpha_c \nabla_{\boldsymbol{w}} \mathcal{L}^c(\boldsymbol{w}). 
	\label{eq-critic-update}
\end{align}
where $\alpha_c$ is the critic learning rate.

For clarity, we write the training principles for both the actor and critic networks in the PPO model in a unified form. Let $\boldsymbol{\theta}=\left[ \boldsymbol{\phi},\boldsymbol{w} \right]$ be the trainable weights of the PPO model. Let $\mathcal{L}(\boldsymbol{\theta})$ be a unified loss function, given by
\begin{align}
	\mathcal{L}(\boldsymbol{\theta}) = -\mathcal{J}^a(\boldsymbol{\phi}) + \mathcal{L}^c(\boldsymbol{w}). 
	\label{eq-unified-loss}
\end{align}
Then, the PPO model update can be written in a unified form, equivalent to~\eqref{eq-actor-update} and~\eqref{eq-critic-update}, as
\begin{align}
	\boldsymbol{\theta} \leftarrow \boldsymbol{\theta} - \alpha \nabla_{\boldsymbol{\theta}} \mathcal{L}(\boldsymbol{\theta})
	\label{eq-update-unified}
\end{align}
where $\alpha$ is set to $\alpha_a$ and $\alpha_c$ for training $\boldsymbol{\phi}$ and $\boldsymbol{w}$ respectively.

\subsection{Background of Meta Reinforcement Learning}

Let $\bar{\boldsymbol{\theta}}$ denote the weights of a meta RL model for MDP tasks sampled from $\rho(\mathcal{T})$. 
Let $\boldsymbol{\theta}_i$ denote the weights of an individual RL model for MDP task $\mathcal{T}_i \sim \rho(\mathcal{T})$, and let  $\mathcal{L}_{\mathcal{T}_i}\left(\boldsymbol{\theta}_i\right)$ denote the loss function for training the individual RL model.

Mathematically, the meta model parameters, $\bar{\boldsymbol{\theta}}$, are trained by solving the following optimization problem
\begin{align} 
	\min_{\bar{\boldsymbol{\theta}}} \sum_{\mathcal{T}_i \sim \rho(\mathcal{T})} \mathcal{L}_{\mathcal{T}_i} \left( \bar{\boldsymbol{\theta}} - \alpha \nabla_{\boldsymbol{\theta}} \mathcal{L}_{\mathcal{T}_i}\left(\bar{\boldsymbol{\theta}}\right) \right). 
    \label{eq-meta-obj}
\end{align}
In~\eqref{eq-meta-obj}, $\mathcal{L}_{\mathcal{T}_i} \left( \bar{\boldsymbol{\theta}} - \alpha \nabla_{\boldsymbol{\theta}} \mathcal{L}_{\mathcal{T}_i}\left(\bar{\boldsymbol{\theta}}\right) \right)$ evaluates the loss of the individual model for task $\mathcal{T}_i$ after one gradient descent step from the initial model weights, $\bar{\boldsymbol{\theta}}$, referred to as adaptation loss. 
By minimizing the total adaptation loss of all possible tasks sampled from $\rho(\mathcal{T})$, an optimal meta model can be obtained~\cite{finn2017model,wang2022meta,yuan2021meta}. 
To solve 
\eqref{eq-meta-obj}, a stochastic gradient descent method can be used, to iteratively update $\bar{\boldsymbol{\theta}}$ in multiple meta iterations until convergence. 
Each meta iteration involves one gradient descent over $\bar{\boldsymbol{\theta}}$ based on the meta model's loss function in~\eqref{eq-meta-obj}, referred to as meta update, and given by
\begin{align} 
	\bar{\boldsymbol{\theta}} \leftarrow  \bar{\boldsymbol{\theta}} - \beta  \frac{1}{I}\sum_{i=1}^I  \nabla_{\bar{\boldsymbol{\theta}}}  \mathcal{L}_{\mathcal{T}_i} \left( \bar{\boldsymbol{\theta}} - \alpha \nabla_{\bar{\boldsymbol{\theta}}} \mathcal{L}_{\mathcal{T}_i}\left(\bar{\boldsymbol{\theta}}\right) \right)  
    \label{eq-meta-gradient}
\end{align}
where $\beta$ is the meta step size, and $I$ is the task batch size.
It is time consuming to directly compute~\eqref{eq-meta-gradient} by using deep learning libraries such as Tensorflow, as it involves the calculation of Hessian matrix. First-order approximation methods have been proposed to simplify the calculation of~\eqref{eq-meta-gradient} without remarkable errors~\cite{nichol2018first}.

\subsection{Meta PPO Algorithm}

\begin{algorithm}[t]
\DontPrintSemicolon
\SetKwInOut{Input}{Input}{}
\SetKwInOut{Output}{Output}
\SetKwInput{Define}{Define}
\SetKwInput{Let}{Let}
\SetKwInput{Find}{Find}
\tcc{Meta Training Phase}
Randomly initialize $\bar{\boldsymbol{\theta}}=\left[ \bar{\boldsymbol{\phi}},\bar{\boldsymbol{w}} \right]$; \\
\For{each meta iteration }{ 
    \For{$1 \leq i \leq I$}{ 
    	Sample a random MDP task $\mathcal{T}_i \sim \rho(\mathcal{T})$; \\
        Initialize individual PPO model weights as $\bar{\boldsymbol{\theta}}$; \\
        Collect a trajectory set, $\mathcal{D}_i = \left\{ \tau_i^{\bar{\boldsymbol{\theta}}} \right\}$; \\
        Individual model update based on $\mathcal{D}_i$: $\tilde{\boldsymbol{\theta}}_i \leftarrow \bar{\boldsymbol{\theta}} - \alpha \nabla_{\bar{\boldsymbol{\theta}}} \mathcal{L}_{\mathcal{T}_i}\left(\bar{\boldsymbol{\theta}}\right)$; \\
        Use the adapted PPO model with weights $\tilde{\boldsymbol{\theta}}_i$ to collect a new trajectory set: $\mathcal{D}_i' = \left\{ \tau_i^{\tilde{\boldsymbol{\theta}}_i}\right\}$; \\
        Calculate adaptation loss gradient $\nabla_{\bar{\boldsymbol{\theta}}}  \mathcal{L}_{\mathcal{T}_i} \left( \bar{\boldsymbol{\theta}} - \alpha \nabla_{\bar{\boldsymbol{\theta}}} \mathcal{L}_{\mathcal{T}_i}\left(\bar{\boldsymbol{\theta}}\right) \right)$  based on $\mathcal{D}_i'$; \\
    }       
    \textbf{Meta update}: Gradient descent based on~\eqref{eq-meta-gradient}. \\
}   
\tcc{Meta Adaptation Phase}
Given trained meta model weights $\bar{\boldsymbol{\theta}}=\left[ \bar{\boldsymbol{\phi}},\bar{\boldsymbol{w}} \right]$; \\
\For{any unseen task $\mathcal{T}_{i'} \sim \rho(\mathcal{T})$}{ 
Initialize individual PPO model weights $\boldsymbol{\theta}_{i'} = \bar{\boldsymbol{\theta}}$; \\
Perform gradient descent, $ \boldsymbol{\theta}_{i'} \leftarrow \boldsymbol{\theta}_{i'}  - \alpha \nabla_{\boldsymbol{\theta}_{i'}} \mathcal{L}_{\mathcal{T}_i}\left(\boldsymbol{\theta}_{i'}\right)$, for several steps.
}
\caption{Meta PPO Algorithm}
\label{alg:Meta-PPO}
\end{algorithm}

Let $\bar{\boldsymbol{\theta}}=\left[ \bar{\boldsymbol{\phi}},\bar{\boldsymbol{w}} \right]$ denote the weights of the meta PPO model, and let $\boldsymbol{\theta}_i=\left[ \boldsymbol{\phi}_i,\boldsymbol{w}_i \right]$ denote the weights of the individual PPO
model for MDP task $\mathcal{T}_i \sim \rho(\mathcal{T})$.
The meta PPO algorithm includes two phases: \emph{meta training phase} and \emph{meta adaptation phase}, with pseudo codes given in Algorithm~\ref{alg:Meta-PPO}.

In each meta training iteration, a number, $I$, of MDP tasks are sampled, and the following steps are performed between a meta learning agent and $I$ PPO agents.
\begin{itemize}

    \item \textbf{Individual model initialization:} The meta learning agent distributes the current meta model to the $I$ PPO agents, and the PPO model for task $\mathcal{T}_i$ is initialized with weights $\boldsymbol{\theta}_i=\bar{\boldsymbol{\theta}}=\left[ \bar{\boldsymbol{\phi}},\bar{\boldsymbol{w}} \right]$ (\emph{line} 5);

    \item \textbf{First-round data collection:} Each PPO agent collects a set of trajectory data, $\mathcal{D}_i = \left\{ \tau_i^{\bar{\boldsymbol{\theta}}} \right\}$, where $\tau_i^{\bar{\boldsymbol{\theta}}}=\{\boldsymbol{s}_1, \boldsymbol{a}_1, r_1, \cdots, \boldsymbol{s}_T, \boldsymbol{a}_T, r_T\}$ denotes a trajectory of state-action-reward transitions over $T$ time steps. Here, the superscript in $\tau_i^{\bar{\boldsymbol{\theta}}}$ indicates that the data are collected through the interaction with the network environment by using model weights $\bar{\boldsymbol{\theta}}$ (\emph{line} 6);

    \item \textbf{Individual model adaptation:} With trajectory data $\mathcal{D}_i$, each PPO agent $i$ performs a gradient descent to obtain adapted PPO model weights, $\tilde{\boldsymbol{\theta}}_i \leftarrow \bar{\boldsymbol{\theta}} - \alpha \nabla_{\bar{\boldsymbol{\theta}}} \mathcal{L}_{\mathcal{T}_i}\left(\bar{\boldsymbol{\theta}}\right)$, according to~\eqref{eq-update-unified} (\emph{line} 7);

    \item \textbf{Second-round data collection:} An extra trajectory set, $\mathcal{D}_i' = \left\{ \tau_i^{\tilde{\boldsymbol{\theta}}_i}\right\}$, is collected at each PPO agent by using the adapted model weights, $\tilde{\boldsymbol{\theta}}_i$ (\emph{line} 8);

    \item \textbf{Adaptation loss evaluation:} Based on $\mathcal{D}_i'$, the adaptation loss gradients, $\nabla_{\bar{\boldsymbol{\theta}}}  \mathcal{L}_{\mathcal{T}_i} \left( \bar{\boldsymbol{\theta}} - \alpha \nabla_{\bar{\boldsymbol{\theta}}} \mathcal{L}_{\mathcal{T}_i}\left(\bar{\boldsymbol{\theta}}\right) \right)$, are computed and sent to the meta learning agent (\emph{line} 9);

    \item \textbf{Meta model update:} After obtaining the adaptation loss gradients of $I$ tasks, the meta learning agent updates the meta model weights, $\bar{\boldsymbol{\theta}}$, according to~\eqref{eq-meta-gradient} (\emph{line} 10).

\end{itemize}

The meta model is iteratively trained until it captures the general features of all possible tasks sampled from $\rho(\mathcal{T})$, indicated by the convergence of the total adaptation loss in~\eqref{eq-meta-obj}.
In the meta adaptation phase, a PPO model for any unseen task $\mathcal{T}_{i'} \sim \rho(\mathcal{T})$ during the meta training process can be quickly adapted from the trained meta model, via several gradient descent steps based on~\eqref{eq-update-unified} (\emph{lines} 13 and 14). 

\section{Simulation Results}
\label{sec:Simulation Results}

\subsection{Simulation Setup}

We consider a moving vehicle cluster on a four-lane unidirectional highway segment, which consists of $3$ CAV pairs and $10$ other vehicle users such as human-driven vehicles (HDV). The delay requirement for object classification is $\Delta = 100$ $ms$. In each time slot, each HDV has a transmission request and occupies a bandwidth of $0.5$ MHz according to a $Bernoulli(0.5)$ distribution. Consider a transmission priority for HDVs, as the CAVs can work in the default SP mode without transmission requirements. With a total radio spectrum bandwidth of $B = 10.5$ MHz, the available bandwidth for the CAVs at time slot $t$ is $B(t) = B-0.5M(t)$, where $M(t)$ is
the number of HDVs that have transmission requests.

For CAV pair $k$, the transitions of shared workload, $W_k(t)$, across different time slots follow a Markov chain with states in
$\{4, 5, 6, 7, 8\}$. For a stationary network environment, the state transition probability matrix for the workload is deterministic.
To simulate nonstationary network environments, we consider diverse MDP tasks with different state transition probability matrices for the workload. Specifically, we consider four categories of MDP tasks, with the average steady-state probabilities for the workload given in Table~\ref{Table:Steady-state probabilities of transition matrix}, indicating low, medium, high, and general workload environments. A meta PPO model is trained for each category. In comparison with the meta model associated with the general workload category, the first three meta models have a higher customization level.

We implement the meta PPO algorithm in Python 3.7.16. The learning modules are implemented using TensorFlow 2.11.0. The actor and critic networks have two hidden layers with $(16, 16)$ and $(8, 8)$ neurons respectively, both with Relu activation functions. There are $K$ heads associated with the $K$ CAV pairs at the actor’s output layer, each with a two-dimensional output with \texttt{SoftMax} activation, indicating a categorical distribution between CP and SP modes.

\begin{table}[t]
\small
\centering
\caption{\scshape{Nonstationary Network Environment Setup}}
\begin{tabular}{ c | c } 
\hline\noalign{\vskip 0.3mm}\hline\noalign{\smallskip}
\textbf{Workload} & Average steady-state probabilities \\
\noalign{\smallskip}\hline\noalign{\smallskip}
Low &  $\left[0.735, 0.064, 0.060, 0.052, 0.086\right]$ \tabularnewline\noalign{\smallskip}\hline\noalign{\smallskip}
Medium & $\left[0.061, 0.058, 0.775, 0.053, 0.053\right]$ \tabularnewline\noalign{\smallskip}\hline\noalign{\smallskip} 
High & $\left[0.053, 0.132, 0.093, 0.052, 0.670\right]$ \tabularnewline\noalign{\smallskip}\hline\noalign{\smallskip}
General & $\left[0.152, 0.223, 0.247, 0.223, 0.152\right]$ \tabularnewline\noalign{\smallskip}
\hline\noalign{\vskip 0.3mm}\hline\noalign{\smallskip}
\end{tabular}
\label{Table:Steady-state probabilities of transition matrix}
\end{table}

\subsection{Performance Evaluation}

\begin{figure}
    \centering

    \includegraphics[width=1\linewidth]{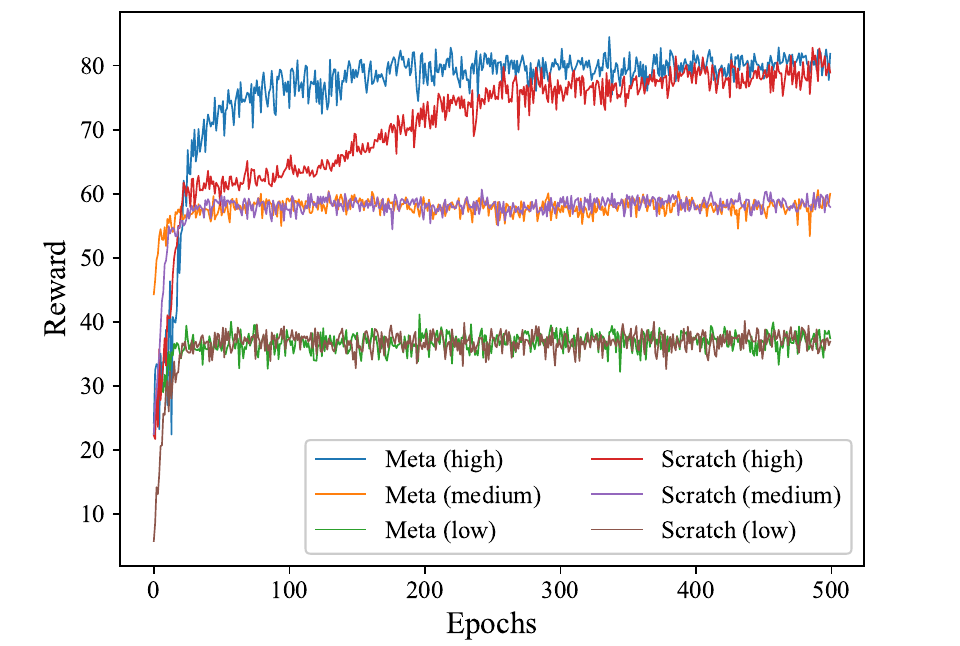} 

  \caption{Performance comparison between training from scratch and meta
adaptation under different workloads.}\label{fig:comp-meta-scratch}
\end{figure}

We first conduct simulations to evaluate how meta learning accelerates the model adaptation in comparison with training
from scratch. We sample one MDP task from each of the low, medium, high workload categories. For each task, two PPO models are trained. One is trained from scratch, and the other is trained with the meta model of corresponding category as the initial point. All the PPO models are trained over $500$ epochs, each consisting of $750$ time steps.
Fig.~\ref{fig:comp-meta-scratch} shows the average reward during the training process of different PPO models. We observe that, for each workload category, the meta model adapts quickly and accurately, as indicated by the faster convergence and the comparable reward
after convergence in comparison with training from scratch. The convergence speed difference between meta adaptation and training from scratch is more significant for high workload environments, demonstrating the superiority of meta learning in more difficult MDP tasks. Moreover, we observe a higher reward after convergence for a higher workload, as the computing demand reduction in the CP mode is proportional to the shared workload, improving the computing efficiency gain. Note that such a trend is valid only with sufficient radio resources~\cite{10413956}.

\begin{figure}
    \centering

    \includegraphics[width=1\linewidth]{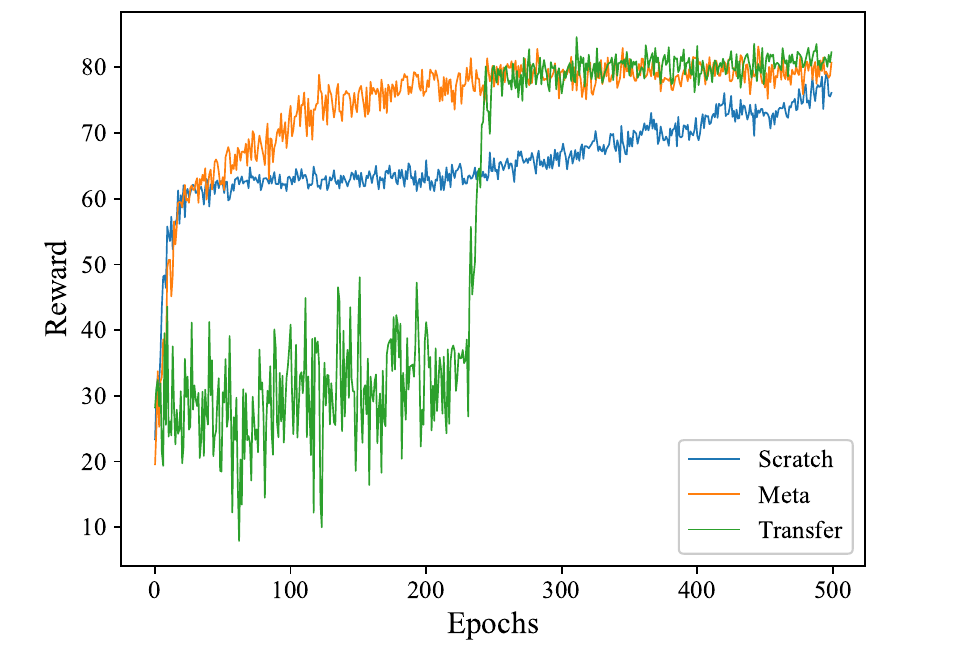}

  \caption{Performance comparison between training from scratch, transfer learning, and meta learning for one task.}\label{fig:PPO_meta_tl}
\end{figure}

Then, we compare the performance of meta learning with a benchmark algorithm, transfer learning, which is another typical algorithm that uses learned knowledge to accelerate model training. We sample two MDP tasks from the general and high workload categories, referred to as task 1 and task 2 respectively. A PPO model is first trained from scratch for task 2. Then, three PPO models are trained for task 1, corresponding to training from scratch, meta model adaption, and transfer learning from task 2. 
Fig.~\ref{fig:PPO_meta_tl} shows the average reward over training epochs for the three PPO models of task 1. We see that meta learning shows the fastest convergence, followed by transfer learning which converges after approximately $300$ epochs. The initial reward of transfer learning is low, as the model is initially customized to task 2. However, due to the shared features between tasks 1 and 2, transfer learning, with some learned knowledge from task 2, converges faster than training from scratch for task 1.

\begin{figure}
    \centering

    \includegraphics[width=1\linewidth]{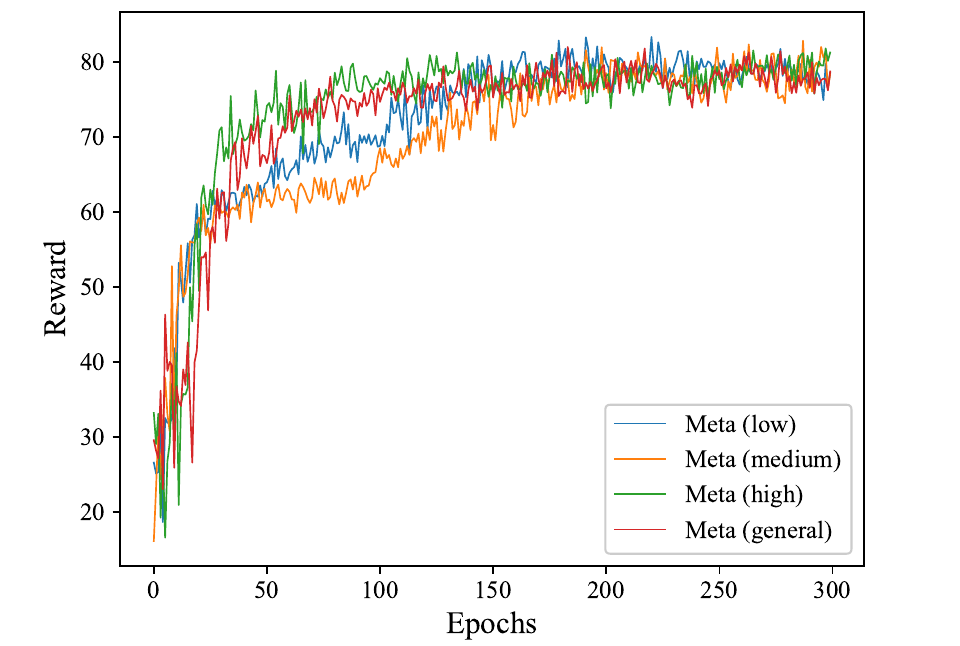}

  \caption{Model adaptation performance with meta models of different customization levels.}\label{fig:customization}
\end{figure}

Finally, we evaluate the impact of the customization levels of meta models on model adaptation performance. We use task 2 as an example. Specifically, with the meta models of the four categories as an initial point, four PPO models are trained for task 2. 
Fig.~\ref{fig:customization} shows that the meta model customized for the high workload category, which matches task 2, achieves
the best performance in terms of the convergence speed. In comparison, the meta models customized for the low and medium workload categories show inferior performance. However, the meta model for the general workload category shows a medium convergence speed, which implies that a general meta model fits a specific task better than meta models customized for other categories.

\section{Conclusion}

In this paper, we consider the nonstationary characteristics in vehicular networks, and develop a meta reinforcement learning based adaptive cooperative perception solution that facilitates fast model adaption based on a meta model with good generalization ability. Simulation results demonstrate the superiority of meta learning for accelerating model adaptation in comparison with training from scratch and transfer learning. We have also obtained useful principles in selecting meta models with different customization levels as initial training points. For future work, we will consider more aspects of the network nonstationarity, such as a varying vehicle cluster size, which potentially requires the integration of graph neural networks to the meta reinforcement learning framework.

\bibliographystyle{IEEEtran}
\bibliography{Ref}

\end{document}